\documentclass[12pt]{article}
\usepackage{latexsym}
\input amssym.def
\input amssym.tex
\makeatletter

\@addtoreset{equation}{section}
\makeatother
\def\tr{\mbox{tr}}

\def\L{{\cal L}}
\def\S{{\cal S}}
\def\half{{\textstyle{\frac{1}{2}}}}

\begin{document}
\begin{titlepage}
\title{\vskip -60pt
{\small
\begin{flushright}
KIAS-P01034\\ hep-th/0108145
\end{flushright}}
\vskip 45pt  On  a Matrix Model of Level Structure} \vspace{4.0cm}
\author{\\
\\
\\Jeong-Hyuck Park\,\thanks{Electronic correspondence: jhp@kias.re.kr}}
\date{}
\maketitle
\vspace{-1.0cm}
\begin{center}
\textit{Korea Institute for Advanced Study}\\ \textit{207-43 Cheongryangri-dong, Dongdaemun-gu}\\ \textit{Seoul
130-012, Korea}
\end{center}
\vspace{2.0cm}
\begin{abstract}
\noindent  We generalize the   dimensionally reduced Yang-Mills matrix model by  adding  $d=1$ Chern-Simons term
and  terms for a bosonic vector. The coefficient, $\kappa$ of the Chern-Simons term must be integer, and hence the
level structure.  We show at the bottom of the Yang-Mills potential, the low energy limit,  only the linear motion
is allowed for D0 particles. Namely  all the particles align themselves on a single straight line subject to
$\kappa^{2}/r^2$ repulsive potential from each other. We argue the relevant brane configuration to be  D0-branes
in a D4 after $\kappa$ of D8's  pass the system.
%%%%
%%%The repulsive potential is  identified as an electric potential in D4 theories
%%%due to the charge acquirement of D0
%%%particles from the $\kappa$ of fundamental string ends generated
%%%by the  Hanany-Witten effect.
%%%%
\end{abstract}

\thispagestyle{empty}
\end{titlepage}
\newpage

%%%%
%%\noindent
%%%
%%The repulsive potential is  identified as an electric potential in
%%D4 theories due to the charge acquirement of D0 particles from
%%the $\kappa$ of fundamental string ends generated
%%by the  Hanany-Witten effect.
%%%
%%%\pacs{PACS numbers : 11.25.-w, 11.27.+d, 02.10.Yn}]
%%\newpage
%%%%

%%%%%%%%%%%%%%%%%%%%%%%%%%%%%%%%%%%%%%%%%%%%%%%%%%%%%%%%%%%%%%%%%%%%%%%%%%%%%%%%%%%%%%%%%%%%%%%%%%%%%%%%%%%%%%%%%%%%%%%
%%%%%%%%%%%%%%%%%%%%%%%%%%%%%%%%%%%%%%%%%%%%%%%%%%%%%%%%%%%%%%%%%%%%%%%%%%%%%%%%%%%%%%%%%%%%%%%%%%%%%%%%%%%%%%%%%%%%%%%
%%%%%%%%%%%%%%%%%%%%%%%%%%%%%%%%%%%%%%%%%%%%%%%%%%%%%%%%%%%%%%%%%%%%%%%%%%%%%%%%%%%%%%%%%%%%%%%%%%%%%%%%%%%%%%%%%%%%%%%
\section{Introduction}
In string theory $N\times N$ matrix models describe the dynamics of $N$ D0-branes.  In particular, according to
the renowned conjecture by Banks, Fischler, Shenker and Susskind \cite{BFSS},  the infinite momentum limit of
M-theory is exactly described by the large $N$ limit of   U($N$) Yang-Mills matrix quantum mechanics.  In the
limit only degrees of freedom are D0 particles and other higher dimensional extended objects appear as composites
of D0-branes. In fact, the dimensionally reduced Yang-Mills  action can be regarded as the matrix regularization
of the theory of membrane  which is the fundamental object in M-theory \cite{HoppePhD}. On the other hand,
considering finite number of D0 particles with various extended objects as background in type IIA string theory
may require more generalized form of the matrix action. \newline

An interesting phenomenon for D0 particles near a D8-brane is that, every time  the D0 particle crosses the
D8-brane, a fundamental string is created connecting  those two D-branes \cite{9705084}. This  is in fact U-dual
to the Hanany-Witten effect \cite{9611230}. For $\kappa$ of  D8-branes there appear $\kappa$ number of fundamental
strings and this induces a $d=1$ Chern-Simons term of level $\kappa$ in the D0-brane action
\begin{equation}
\kappa\,\tr A_{0}\,.
\end{equation}
%%%
%For $N$ D0 particles $A_{0}$ is  $N\times N$ matrix valued.
%%%
This Chern-Simons term can be also regarded as  a chemical potential for string ends \cite{0107200}.\newline

However adding this term alone in  the  dimensionally reduced Yang-Mills action causes a problem. In general, at
quantum level, the Gauss constraint generates unitary transformations,
$U=e^{i\Lambda},\,\Lambda^{\dagger}=\Lambda$ on all the arguments in the wavefunction \cite{0101029}
\begin{equation}
|\Psi^{\prime}\rangle=e^{i\kappa\tr\Lambda}|\Psi\rangle\,.
\end{equation}
Taking the particular choice, $\Lambda=\mbox{diag}(2\pi,0,0,\cdots,0)$  gives the identity matrix, $U=1_{N\times
N}$ and  the Gauss constraint on wavefunctions successfully works only  for integer, $\kappa$, leading the level
quantization as in noncommutative Chern-Simons theories \cite{0102188}. This is also consistent with our
interpretation of $\kappa$ as the number of D8-branes. Now the problem is as follows. As all the arguments are
matrices meaning they are in the adjoint representation, the overall U($1$) transformation will leave the
wavefunction invariant, and this is clearly  inconsistent  with the Gauss constraint for non-zero $\kappa$. In
fact, any SU($N$) singlet wavefunction would be  automatically U($N$) singlet too \cite{9712086}.\newline

Curing the problem above requires the presence of arguments in other representation. In this paper, we consider a
bosonic complex vector, $\psi$ in the fundamental representation  as introduced by Polychronakos  in the context
of quantum Hall physics via a finite matrix Chern-Simons theory \cite{0103013}.  Therefore the action we consider
contains   the dimensionally reduced Yang-Mills terms, $d=1$ Chern-Simons term of level $\kappa$ and  terms for
the vector
\begin{equation}
\S=\displaystyle{\int{\rm
d}t}\,\tr\left(\half\textstyle{\frac{1}{R}}D_{t}X^{I}D_{t}X^{I}-\textstyle{\frac{1}{4}\frac{R\,}{l_{p}^{6}}}
[X^{I},X^{J}]^{2}+\kappa A_{0}+iD_{t}\psi\psi^{\dagger}-V(\psi\psi^{\dagger})\right)\,,
\label{action}
\end{equation}
where $1\leq I,J\leq D$ are space indices, $R$ is the compactification radius and $l_{p}$ is the  Planck length.
As we are mainly interested in the low energy dynamics in this paper,  the kinetic term for $\psi$ is taken here
as nonrelativistic in the sense that it is linear in time derivative as in the Jackiw-Pi model for the
nonrelativistic Chern-Simons vortices \cite{JackiwPi}. In the last section we will be back to this point. \newline

To realize the relevant brane configuration, one needs a D4-brane, as the light D0-D4 string has both fermionic
and bosonic modes while   the   D0-D8 string is fermionic only \cite{9712086,polchinski}. The argument is due to
Susskind and Hellerman \cite{Lenny,0107200}.  Consider D0-branes in a D4-brane and  let D8-branes move far away
after passing the D0-D4 system. The  D0-D8 fundamental strings arising from the Hanany-Witten effect jump one ends
from D8's to the D4 forming D0-D4 strings, and this allows bosonic modes for the strings. We may then regard
$\psi$ as the effective  bosonic degrees. In this picture we put $D=4$.\newline

The covariant derivatives are in our convention
\begin{equation}
\begin{array}{cc}
D_{t}X=\frac{d\,}{dt}X+i[A_{0},X]\,,~&~D_{t}\psi=\frac{d\,}{dt}\psi+iA_{0}\psi\,,
\end{array}
\end{equation}
and the U($N$) gauge symmetry is given by
\begin{equation}
\begin{array}{ccc}
X\rightarrow UXU^{-1},&\psi\rightarrow U\psi,&A_{0}\rightarrow UA_{0}U^{-1}+i\frac{d\,}{dt}UU^{-1}.
\end{array}
\end{equation}

The Hamiltonian is  with $P^{I}=\frac{1}{R}D_{t}X^{I}$
\begin{equation}
H=\tr{\left(\textstyle{\frac{R}{2}}P^{I}P^{I} +\textstyle{\frac{1}{4}\frac{R\,}{l_{p}^{6}}}[X^{I},X^{J}]^{2}
+V(\psi\psi^{\dagger})\right)}\,.
\label{H}
\end{equation}
%%%
%%so that the bosonic vector does not contribute implying  the string vibration freezes.\newline
%%%
The classical equations of motion are
\begin{equation}
\begin{array}{cc}
D_{t}D_{t}X^{I}=-\textstyle{\frac{R^{2}}{l_{p}^{6}}}[[X^{I},X^{J}],X^{J}]\,,&~iD_{t}\psi=V^{\prime}(\psi^{\dagger}\psi)\psi\,,
\end{array}
\label{em}
\end{equation}
and the Gauss constraint is
\begin{equation}
-i[X^{I},P^{I}]+\psi\psi^{\dagger}=\kappa\,1_{N\times N}\,. \label{Gauss}
\end{equation}
Note that without $\psi$ the Gauss constraint would be  problematic at classical level too, since taking trace of
it would require $\kappa=0$. In fact this is the original reason  $\psi$ was introduced in \cite{0103013}. Quantum
mechanically it follows that the general wavefunction  satisfying the Gauss constraint is of the form (see e.g.
\cite{0103179})
\begin{equation}
|\Psi\rangle=S\left(\tr [g_{m}(\bar{c})]\right)\,\displaystyle{\prod_{l=1}^{\kappa}\epsilon^{i_{1}\cdots
i_{N}}(\psi^{\dagger}f_{l_{1}}(\bar{c}))_{i_{1}}\cdots(\psi^{\dagger}f_{l_{N}}(\bar{c}))_{i_{N}}}|0\rangle\,,
\label{wvf}
\end{equation}
where $g_{m},f_{l}$ are arbitrary functions depending on $D$ variables,
$\bar{c}^{I}=\frac{1}{\sqrt{2}}(\frac{1}{R}X^{I}-i{\scriptstyle R}P^{I})$ and $S\left(\tr [g_{m}(\bar{c})]\right)$
is the U($N$) singlet part.\newline

In the next section  we solve the equations of motion and the Gauss constraint classically at the bottom of the
Yang-Mills potential. Physically this corresponds to the low energy limit, $l_{p}\rightarrow 0$,
where D0 particles acquire well defined positions since all the $X^{I}$'s are simultaneously diagonalizable. We show only
the linear motion is allowed for D0 particles.  Namely all the particles align themselves on a single straight
line subject to ${\scriptstyle R}\kappa^{2}/r^2$ repulsive potential from each other.

%%%%%%%%%%%%%%%%%%%%%%%%%%%%%%%%%%%%%%%%%%%%%%%%%%%%%%%%%%%%%%%%%%%%%%%%%%%%%%%%%%%%%%%%%%%%%%%%%%%%%%%%%%%%%%%%%%%%%%%
%%%%%%%%%%%%%%%%%%%%%%%%%%%%%%%%%%%%%%%%%%%%%%%%%%%%%%%%%%%%%%%%%%%%%%%%%%%%%%%%%%%%%%%%%%%%%%%%%%%%%%%%%%%%%%%%%%%%%%%
%%%%%%%%%%%%%%%%%%%%%%%%%%%%%%%%%%%%%%%%%%%%%%%%%%%%%%%%%%%%%%%%%%%%%%%%%%%%%%%%%%%%%%%%%%%%%%%%%%%%%%%%%%%%%%%%%%%%%%%
\section{Solving the Low Energy Dynamics}
Before we solve the dynamics  at the bottom of the Yang-Mills potential, we need to specify carefully what limits
we are looking at. Converting the Hamiltonian into the length scale, $H=1/l_{E}$  we get from Eq.(\ref{H})
$\tr[X,X]^{2}\sim l_{p}^{6}{/(Rl_{E})}$ or
\begin{equation}
[X,X]\sim\frac{l_{p}^{3}}{\sqrt{Rl_{E}}}\,M\,, \label{com}
\end{equation}
where $M$ is a dimensionless matrix of order ``one''. Hence with the unit length, $l_{unit}$   the following limit
vanishes the commutator
\begin{equation}
\frac{l_{p}^{3}}{\sqrt{Rl_{E}}}<<l_{unit}^{2}\,.
\label{limit1}
\end{equation}
Substituting Eq.(\ref{com}) into the equation of motion~(\ref{em}) we get
\begin{equation}
D_{t}D_{t}X\sim\frac{R^{3/2}}{l_{p}^{3}\,l_{E}^{1/2}}[M,X]\,.
\end{equation}
Thus in the limit
\begin{equation}
\frac{R^{3/2}}{l_{p}^{3}\,l_{E}^{1/2}}<<\frac{1}{l_{unit}^{2}}
\label{limit2}
\end{equation}
the equation of motion  becomes that of free motion.  All together, from Eqs.(\ref{limit1},\ref{limit2}) we
require
\begin{equation}
\frac{R^{3}l_{unit}^{4}}{l_{E}}<<l_{p}^{6}<<Rl_{E}l_{unit}^{4}\,.
\end{equation}
This  limit can be realized  as taking $R$ constant,  $l_{p}$ small and $l_{E}$ very large. For example, with a dimensionless  small parameter, $\varepsilon<< 1$,  $l_{p}=\varepsilon\,l_{unit}$, $l_{E}=\varepsilon^{-7}l_{unit}$.\newline
%%%%
%%%In this set up we get
%%%\begin{equation}
%%%R^{3}l_{unit}^{3}\lambda^{7}<<\lambda^{6} l_{unit}^{6}<<R\lambda^{-7}l_{unit}^{5}\,.
%%%\end{equation}
%%%Taking $\lambda\rightarrow 0$ limit  this equation holds.
%%%This corresponds to a ``particular'' low energy limit.
%%%%

At the bottom of the potential, $[X^{I},X^{J}]=0$, all the $X^{I}$'s are simultaneously diagonalizable and we do
so using the gauge symmetry, $X^{I}=\mbox{diag}(x^{I}_{1},x^{I}_{2},\cdots,x^{I}_{N})$. Furthermore the remaining
$\mbox{U}(1)^{N}$ symmetry enables us to set the vector  be real and nonnegative, $\psi=\psi^{\ast}\geq 0$. Now
the $(a,b)$ component of the Gauss constraint is  of the form
\begin{equation}
-\textstyle{\frac{1}{R}}(x_{a}-x_{b})^{2}A_{0ab}+\psi_{a}\psi_{b}=\kappa\,\delta_{ab}\,,
\end{equation}
which determines $\psi$ and the off-diagonal component of $A_{0}$
\begin{equation}
\begin{array}{cc}
\psi_{a}=\sqrt{\kappa}\,,~~&~~~A_{0ab}=\displaystyle{\frac{\kappa{\scriptstyle
R}}{(x_{a}-x_{b})^{2}}}~~\mbox{for~~}a\neq b\,.
\end{array}
\end{equation}
The diagonal element of $A_{0}$ is given by the equation of the motion for $\psi$ (\ref{em})
\begin{equation}
\begin{array}{cc}
V^{\prime}(\kappa N)+\displaystyle{\sum_{b}}\,A_{0ab}=0\,,~~&~~A_{0aa}=-\displaystyle{\sum_{b\neq
a}}\,\displaystyle{\frac{\kappa{\scriptstyle R}}{(x_{a}-x_{b})^{2}}}-V^{\prime}(\kappa N)\,.
\end{array}
\end{equation}
Note that  $\psi$ freezes becoming non-dynamical   and $A_{0}$ is real and symmetric.\newline

The remaining main equation of motion at the bottom of the potential is
\begin{equation}
0=D_{t}D_{t}X^{I}=\left(\textstyle{\frac{d^{2}\,}{dt^{2}}}X^{I}-[A_{0},[A_{0},X^{I}]]\right)+i\left(
[\textstyle{\frac{d\,}{dt}}A_{0},X^{I}]+2[A_{0},\textstyle{\frac{d\,}{dt}}X^{I}]\right)\,.
\end{equation}
Here the formula is written  as a sum of  symmetric part and anti-symmetric part, and we require both to vanish separately. 
%%
%\begin{equation}
%\begin{array}{cc}
%\textstyle{\frac{d^{2}\,}{dt^{2}}}X^{I}-[A_{0},[A_{0},X^{I}]]=0\,,~~~&~~~
%[\textstyle{\frac{d\,}{dt}}A_{0},X^{I}]+2[A_{0},\textstyle{\frac{d\,}{dt}}X^{I}]=0\,.
%\end{array}
%\end{equation}
%%
Using
$\frac{d\,}{dt}A_{0ab}=-2A_{0ab}(x_{a}^{J}-x_{b}^{J})(\frac{d\,}{dt}x_{a}^{J}-\frac{d\,}{dt}x_{b}^{J})/(x_{a}-x_{b})^{2}$
for $a\neq b$, the anti-symmetric part reads
\begin{equation}
\displaystyle{\left(\delta^{IJ}-\frac{(x_{a}^{I}-x_{b}^{I})(x_{a}^{J}-x_{b}^{J})}{(x_{a}-x_{b})^{2}}\right)}
(\textstyle{\frac{d\,}{dt}x_{a}^{J}-\frac{d\,}{dt}x_{b}^{J}})=0\,,
\end{equation}
which says simply $\vec{x}_{a}-\vec{x}_{b}$ is parallel to $\frac{d\,}{dt}\vec{x}_{a}-\frac{d\,}{dt}\vec{x}_{b}$. On the other hand,   
%%%
%%In particular this implies from
%%\begin{equation}
%%0=\textstyle{(\frac{d\,}{dt}\vec{x}_{a}-\frac{d\,}{dt}\vec{x}_{b})
%%+(\frac{d\,}{dt}\vec{x}_{b}-\frac{d\,}{dt}\vec{x}_{c})
%%+(\frac{d\,}{dt}\vec{x}_{c}-\frac{d\,}{dt}\vec{x}_{a})}
%%\end{equation}
%%
%%that $\vec{x}_{a},\vec{x}_{b},,\vec{x}_{c}$ are on a single plane or 
%%all the D0 particles move on a plane. This
%%will be further restricted to a linear motion shortly.\newline
%%%
the  off-diagonal component of the symmetric part gives  the key formula to solve the system. With $a\neq b$
\begin{equation}
(\vec{x}_{a}-\vec{x}_{b})A_{0ab}(A_{0aa}-A_{0bb}) =\displaystyle{\sum_{c\neq a,b}\left((\vec{x}_{a}-\vec{x}_{c})
+(\vec{x}_{b}-\vec{x}_{c})\right)A_{0ac}A_{0bc}}\,.
\label{key}
\end{equation}
To see the geometric meaning we consider  a $(D-1)$-dimensional hyperplane with a unit vector, $\hat{u}$ 
\begin{equation}
\hat{u}\cdot\vec{x}=l\geq 0\,.
\end{equation}
Choosing the direction, $\hat{u}$ and the length, $l$ properly one can show that there exists a hyperplane such that it contains at least two points, say $\vec{x}_{a},\vec{x}_{b}$ 
\begin{equation}
\hat{u}\cdot\vec{x}_{a}=\hat{u}\cdot\vec{x}_{b}=l\,,
\end{equation}
and further that for any point, $\vec{x}_{c}$
\begin{equation}
\begin{array}{ccc}
\hat{u}\cdot\vec{x}_{c}\leq l~~&~~\mbox{or}~~&~~\hat{u}\cdot(\vec{x}_{a}-\vec{x}_{c})=\hat{u}\cdot(\vec{x}_{b}-\vec{x}_{c})\geq 0\,.
\end{array}
\end{equation}
We may regard the hyperplane as a ``boundary'' for the points. Now the scalar product of Eq.(\ref{key}) with the unit vector shows from $A_{0ac}A_{bc}>0$ that  all the points must lie on the plane, $\hat{u}\cdot\vec{x}_{c}=l$. We can repeat the argument applying to a smaller dimensional hyperplane until  we end up with the final statement  that  all the points  must lie  on a single straight line! Consequently  the  anti-symmetric part is automatically satisfied\footnote{Similar collinear motion was also observed  in a matrix model with a certain ansatz \cite{9705047}.}. \newline

We set with a $D$-dimensional unit vector, $\hat{n}$
\begin{equation}
\begin{array}{cc}
\vec{x}_{a}=\vec{x}_{CM}+y_{a}\hat{n}\,,~~~~&~~~\displaystyle{\sum_{a}}\,y_{a}=0\,.
\end{array}
\end{equation}
This reduces all the equations of motion to a single equation
\begin{equation}
\textstyle{\frac{d^{2}\,}{dt^{2}}}y_{a}=\displaystyle{\sum_{b\neq
a}\,\frac{2{\scriptstyle{R}}^{2}\kappa^{2}}{(y_{a}-y_{b})^{3}}}\,.
\end{equation}
The center of mass moves with a constant velocity and  the unit vector is time independent. The effective
Lagrangian for this linear motion is given by ${\scriptstyle R}\kappa^{2}/r^2$ repulsive potential
\begin{equation}
\L_{eff} =\textstyle{\half\frac{N}{R}}(\textstyle{\frac{d\,}{dt}}\vec{x}_{CM})^{2}
+\displaystyle{\sum_{a}\,}\half\textstyle{\frac{1}{R}}(\textstyle{\frac{d\,}{dt}}y_{a})^{2}-
\displaystyle{\sum_{a>b}\,\frac{{\scriptstyle{R}}\kappa^{2}}{(y_{a}-y_{b})^{2}}}.
\end{equation}

In particular, for two particle system, $N=2$,  we get the general solution in a closed form
\begin{equation}
\begin{array}{cc}
\vec{x}_{1}=\vec{x}_{CM}+\hat{n}\,\displaystyle{\sqrt{(vt)^{2}+({\scriptstyle{R}}\kappa/2v)^{2}}}\,,~&~\vec{x}_{2}=2\vec{x}_{CM}-\vec{x}_{1}\,,
\end{array}
\end{equation}
where $v$ is an arbitrary constant  corresponding to the eternal  velocity.

%%%%%%%%%%%%%%%%%%%%%%%%%%%%%%%%%%%%%%%%%%%%%%%%%%%%%%%%%%%%%%%%%%%%%%%%%%%%%%%%%%%%%%%%%%%%%%%%%%%%%%%%%%%%%%%%%%%%%%%
%%%%%%%%%%%%%%%%%%%%%%%%%%%%%%%%%%%%%%%%%%%%%%%%%%%%%%%%%%%%%%%%%%%%%%%%%%%%%%%%%%%%%%%%%%%%%%%%%%%%%%%%%%%%%%%%%%%%%%%
%%%%%%%%%%%%%%%%%%%%%%%%%%%%%%%%%%%%%%%%%%%%%%%%%%%%%%%%%%%%%%%%%%%%%%%%%%%%%%%%%%%%%%%%%%%%%%%%%%%%%%%%%%%%%%%%%%%%%%%
\section{Discussion}
We have shown  that at the bottom of the Yang-Mills potential only the linear motion is allowed for D0 particles
subject to ${\scriptstyle R}\kappa^{2}/r^2$ repulsive potential from each other. The bottom of the potential is a
region where the theory is well described by the classical analysis, since it corresponds to the low energy limit
and all the D0 particles acquire well defined positions. Our results tell us  that at low energy D0 particles tend
to align themselves on a straight line and due to the repulsive potential they move far from each other. This
implies they can not form a bound state except $N=1$ case, a single D0 particle. \newline

As a relevant brane configuration we have  proposed a D0-D4-D8 system which was originally conceived by Susskind
in the context of stringy quantum Hall system. Namely D0-branes in a D4 after $\kappa$ of D8's  pass them.  Our
results agree with this picture in several aspects. \newline (i) Classically  $\psi_{a}=\sqrt{\kappa}$ result
shows the number of strings attached to each D0 particle is $\kappa$, as the  number operator for $\psi_{a}$ is
$\psi^{\dagger}_{a}\psi_{a}$. At quantum level we also note  the wavefunction\,(\ref{wvf}) has eigenvalue $\kappa
N$ for the number operator's sum $\sum_{a}\,\psi^{\dagger}_{a}\psi_{a}$. (ii) The  brane configuration where D4 is
not parallel to D8's  breaks all the supersymmetries leading to a non-BPS configuration. This is also consistent
with the presence of the repulsive potential. One subtle case to note is  for the system with a  single D0
particle, as it can be   stable  though it is not BPS. (iii) The $\kappa$ and $r$ dependence in the repulsive
potential is consistent with the  brane configuration.  The electric charge of D0 particles  acquired from the
$\kappa$ of string ends is of  $\kappa$ unit, and hence $\kappa^{2}$ coefficient.  The electric field is confined
in D4, the four-dimensional space, and hence $r^{-2}$ behaviour.\newline

Instead of the nonrelativistic kinetic term for the fundamental strings, we may consider the relativistic form,
$D_{t}\psi D_{t}\psi^{\dagger}$. In this case the kinetic energy of the strings contributes to the Hamiltonian
contrast to the nonrelativistic case, and after quantization $\psi$ generates two harmonic oscillators
corresponding to two different orientations of the  strings. This is analogue to the presence of  particles and
anti-particles  in relativistic field theories. A new aspect of the Gauss constraint is that  $\kappa$ is the
\textit{difference} between the numbers of  those two strings of different orientations ending on each D0
particle. Consequently  the total number of strings attached to a D0-brane is not less than $\kappa$. This shows
at low energy only $\kappa$, the minimum number of strings  end on a D0 particle and the D0-D4-D8 system can be
effectively described by  our action having the nonrelativistic kinetic term.\newline

Finally we comment, apart from the stringy interpretation, that imposing the periodic boundary condition our model
with  $D=1$ choice reduces to the Sutherland model, a  model on a circle with $1/r^{2}$ potential
\cite{Sutherland,PLB26629}.
\newline
\newline
\newline
\begin{center}
\large{\textbf{Acknowledgements}}
\end{center}
The author  wishes to thank D. Bak, J. Hoppe, S. Hyun,  Y. Kiem, H-W Lee, K. Lee,  S. Lee, Y. Michishita, J. Park, A. Polychronakos, 
H. Shin, L. Susskind and P. Yi for stimulating discussions at various stages of writing up this paper.  Part of
this work was carried out during the fruitful PIMS-APCTP summer workshop  at Simon Fraser university.
\newline
\newline

%%%
%%\bibliographystyle{unsrt}
%%\bibliography{reference}
%%%

\end{document}